\documentclass[prb,twocolumn,superscriptaddress]{revtex4-1}
\usepackage{amsthm,amsmath,amssymb,mathrsfs}
\usepackage{color}
\usepackage{graphicx}
\usepackage{dcolumn}
\usepackage{bm}
\usepackage[bookmarks=true,colorlinks,%
            citecolor=blue,linkcolor=blue, 
            breaklinks=true]{hyperref}
\usepackage{setspace}
\usepackage{subfigure}
\usepackage{enumerate}
\usepackage{MnSymbol}

\begin{document}

\title{Ground state phase diagram and the exotic phases in
  the spin-$1/2$ square lattice $J_1$-$J_2$-$J_{\chi}$ model}
\author{Jianwei Yang}
\affiliation{School of Microelectronics and Data Science, Anhui University of Technology, Maanshan 243002, China}
\author{Zhao Liu}
\affiliation{School of Physics, Zhejiang University, Hangzhou
  310058, China}
\author{Ling Wang}
\email{lingwangqs@zju.edu.cn}
\affiliation{School of Physics,
  Zhejiang University, Hangzhou 310058, China}

\begin{abstract}
The intricate interplay between frustration and spin chirality,
induced by the Dzyaloshinskii-Moriya interaction, holds promise for
unveiling novel phases in frustrated quantum magnets. This study
investigates the ground state phase diagram of the spin-$1/2$ square
lattice $J_1$-$J_2$ model upon incorporation of the chiral interaction
$J_{\chi}$. Employing exact diagonalization (ED) with full lattice
symmetries, we analyze the phase evolution as a function of $J_2$ at
fixed $J_{\chi}$, utilizing highly symmetric energy levels. Critical
level crossings and ground state fidelity susceptibility (FS)
techniques aid in pinpointing phase boundaries: Magnetic to
non-magnetic phase transitions are identified through critical level
crossings between gapless magnetic excitations and quasi-degenerate
ground states of non-magnetic phases. Direct transitions between two
non-magnetic phases are characterized by FS peaks due to avoided
ground state level crossing. Based on these observations, we identify
an anticipated chiral spin liquid (CSL) state and an adjacent nematic
spin liquid (NSL) phase within a significant range of $J_{\chi}$,
demarcated by a nearly vertical boundary line at $J_2 \approx
0.65$. This critical line terminates at the lower boundary in
$J_{\chi}$ of a magnetic ordered chiral spin solid (CSS) phase, which
gains prominence with increasing $J_{\chi}$ from both CSL and NSL
phases. The topological nature of the CSL is confirmed using the
modular $\mathcal{S}$ matrix of minimally entangled states (MES) on a
torus and the entanglement spectra (ES) of even and odd sectors on a
cylinder, employing ED and $\rm{SU}(2)$-symmetric density matrix
renormalization group (DMRG) method respectively. Furthermore, a
comprehensive discussion on the nature of the NSL is provided,
exploring aspects such as ground state degeneracy, local bond energy
landscape, and singlet and triplet gaps on various tori, offering
substantial evidence supporting the nematic nature of the NSL.
\end{abstract}

\date{\today}

\maketitle

\section{\label{introduction} Introduction}
Fractional Quantum Hall (FQH) states are among the most exotic phases
of matter in strongly correlated quantum systems which exhibit
fractionalized excitations and gapless edge
modes~\cite{Laughlin83,Haldane88,Moore91}. Interestingly topological
flat bands of fermions or hard-core bosons can host similar physics
without an external magnetic field~\cite{Sheng11,Wang11}. In this
scenario, the role of magnetic flux is emulated by a complex hopping
phase~\cite{Haldane88,Sheng11,Wang11}. In quantum spin system, chiral
spin liquid (CSL) state, akin to the FQH state, is anticipated when
introducing a scalar chirality term ($J_{\chi}$) expressed as
$(\mathbf{S}_i\times\mathbf{S}_j)\cdot\mathbf{S}_k$. The interactions
governed by $J_{\chi}$ break the time-reversal symmetry, playing a
role analogous to that of a magnetic
flux~\cite{bauer_chiral_2014,Nielsen13,Wietek17}. Drawing a parallel
to the electron filling quantized orbitals in a magnetic field, one
can understand spin-up or spin-down states as akin to filled or vacant
states of hard-core bosons. When the intricate interplay between
frustration and chirality occurs, the potential arises for the
generation of novel phases of
matter~\cite{huang_coexistence_2022}. This intersection poses a
significant challenge in condensed matter theory, and demands thorough
exploration and understanding.

In simple geometrically frustrated quantum spin systems, such as
Kagome~\cite{White11,Liao17,He17} and
triangular~\cite{ZhuZY15,Hu_triangulr_15,Saadatmand15} lattice
Heisenberg models, the absence of a spin chirality term $J_{\chi}$
normally results in a ground state that maintains time reversal
symmetry. In these cases, a non-chiral gapped
$\rm{Z}_2$~\cite{White11,ZhuZY15,ZhuZY18,Hu_triangulr_15} or gapless
$\rm{U}(1)$~\cite{Liao17,He17} quantum spin liquid (QSL) is commonly
expected. However, with the introduction of a chiral interaction
$J_{\chi}$ or a plaquette-wise hopping term with complex
phases~\cite{Poilblanc1,Poilblanc2}, topological CSLs can be easily
induced~\cite{bauer_chiral_2014,Wietek17}, since they are
energetically favored compared to the neighboring non-chiral
counterparts~\cite{Wietek17}.

Spontaneous time-reversal symmetry breaking is difficult to detect due
to ground state manifold doubling. It has been observed in both a
triangular lattice Hubbard model~\cite{Szasz20} and its spin version
with four-site interactions in the large U
limit~\cite{Cookmeyer21}. Recently, Sun et al. identified a possible
CSL with this symmetry breaking in a frustrated Kagome spin
model~\cite{SunKagome24}, aided by advanced numerical methods like
Gutzwiller projected parton wavefunctions~\cite{SunKagome24}, which
help distinguish different topological sectors.

For bipartite lattices, similar scenario is anticipated. Specifically,
in the frustrated spin-$1/2$ $J_1$-$J_2$ antiferromagnetic Heisenberg
model on square lattice, without $J_{\chi}$, a gapless QSL has been
identified adjacent to an antiferromagnetic (AFM)
state~\cite{Gong14,Morita15,Wang18,Ferrari20,Nomura20,Schackleton21},
with no reported signs of a CSL. This model closely describes the
magnetic interactions among $\rm{Cu}^{2+}$ within the copper oxide
plane of high-T$_c$ superconducting parent compounds. Recently
experiments have observed an anomalous large thermal Hall signal close
to the AFM phase in undoped and underdoped
cuprate~\cite{Grissonnanche19}. It was interpreted as proximity
effects close to a CSL~\cite{Samajdar19}. Following this rationale, we
undertake a study of the ground state phase diagram of the spin-$1/2$
square lattice $J_1$-$J_2$-$J_{\chi}$ Hamiltonian, examining numerical
support to such a senario. In the literature, it has been proven to be
the local parent Hamiltonian of the $\nu=1/2$ Kalmeyer-Laughlin (KL)
state~\cite{Kalmeyer87}, utilizing conformal field correlators'
relationships~\cite{Nielsen12,Nielsen13}.

In the spin-$1$ square lattice $J_1$-$J_2$-$J_{\chi}$ model,
frustration and chirality create a unique coexistence of non-Abelian
topological order and stripe magnetic order. It is evidenced by the
presence of distinctive features: edge modes in the entanglement
spectra (ES)~\cite{LiES08,QiES12} and non-vanishing stripe magnetic
order in the ground state (of the vacuum
sector)~\cite{huang_coexistence_2022}. This finding corroborates the
richness in phase diagram for the frustrated chiral quantum magnets.

In this work, we study the ground state phase diagram of the
spin-$1/2$ square lattice $J_1$-$J_2$-$J_{\chi}$ Hamiltonian using
exact diagonalization (ED) on small
tori~\cite{Laflorencie04,Noack05,Weisse08,Lauchli11,Sandvik10ed}. ED
provides symmetry quantum numbers for each eigenstate, allowing us to
infer ground state degeneracy and local order. We identify phase
boundaries between magnetic and non-magnetic states through critical
level crossings~\cite{Wang18,Ferrari20,Nomura20,Yang22,Wang22}, marked
by a switch between a gapless magnetic excitation and a nearby
quasi-degenerate non-magnetic ground state as coupling strength
varies. For transitions between non-magnetic phases, we use ground
state fidelity susceptibility (FS) peaks as indicators of critical
points~\cite{GuFidelity10,You11,Wang10}.

As the main results of this study, we detect two new gapped
non-magnetic phases absent from the plain $J_1$-$J_2$ model. For a
intermediate $J_{\chi}$ value, on the relative small $J_2<0.65$ side,
we observe a topological CSL reminiscent of the $\nu=1/2$ KL state. On
the larger $J_2>0.65$ side, we observe a nematic spin liquid (NSL).
This novel NSL exhibits a two-fold ground state degeneracy and a
strong bond anisotropy when placed on a cylinder. In addition, it
possesses a gapped $S=1$ magnetic excitation on a torus.

Chiral topological order is analyzed through the modular $\mathcal{S}$
matrix of minimally entangled states (MES) on a
torus~\cite{zhu_minimal_2013} and the entanglement spectra (ES) of
even and odd sectors on a
cylinder~\cite{Wietek17,zhu_minimal_2013,huang_coexistence_2022},
employing ED and $\rm{SU}(2)$-symmetric density matrix renormalization
group (DMRG) method respectively.

The remainder of this paper is organized as follows. In
Sec.~\ref{phasediagram}, we derive the ground state phase diagram
through critical level crossings and FS methods, utilizing results
obtained from a 32-site fully symmetric periodic cluster.
Sec.~\ref{localorders} focuses on local expectation values,
encompassing various magnetic and valence bond orders across the
entire parameter space. In Sec.~\ref{csl}, we uncover the topological
nature of the CSL, by examining the modular $\mathcal{S}$ matrix from
the MES on a torus, and by the ES on even and odd sectors of a
cylinder. Section~\ref{nsl} focuses on the NSL. We employ the local
bond energy landscape on a cylinder to showcase the two-fold ground
state degeneracy, and compare the ES between the CSL and the NSL,
highlighting their distinctions. We present singlet and triplet gaps
on various tori to confirm the existence of a finite magnetic
gap. Finally, in Section~\ref{conclusion}, we provide a summary along
with discussions and conclusions.

\begin{figure}
\centering
\includegraphics[width=0.5\textwidth]{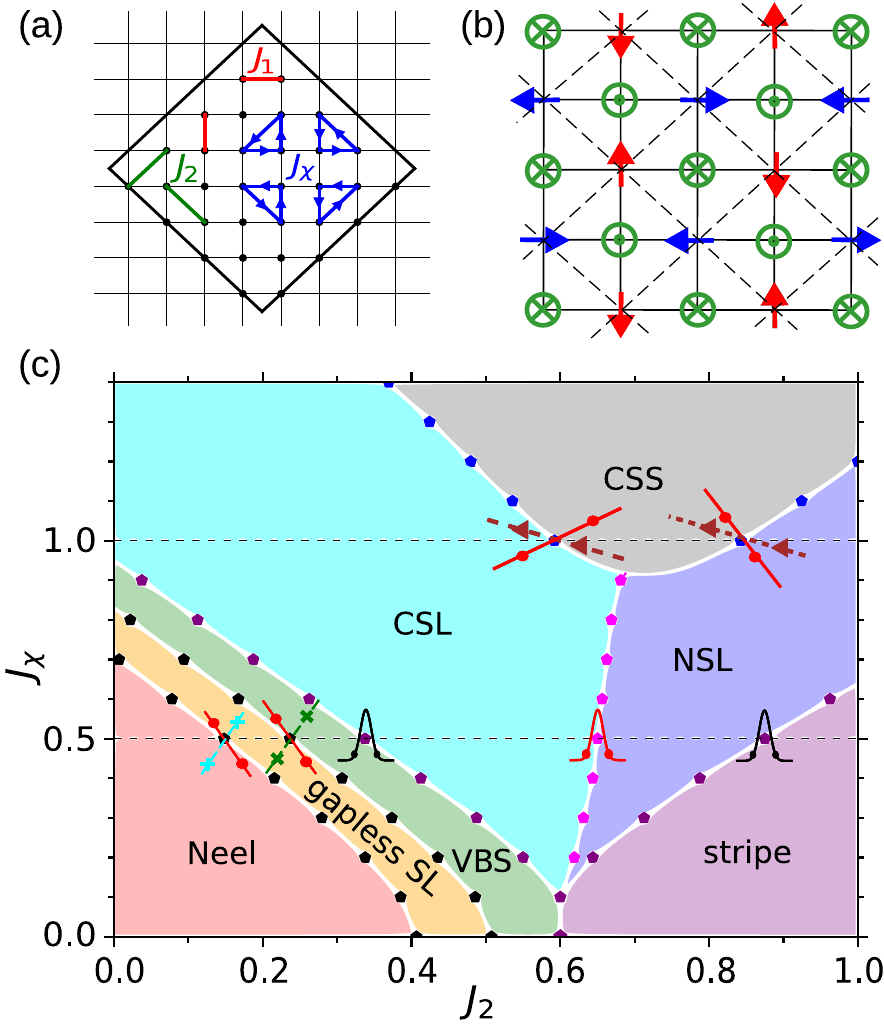}
\caption{(a) Illustration of the three terms in the square lattice
  $J_1$-$J_2$-$J_{\chi}$ Hamiltonian on a 32-site periodic
  cluster. (b) Schematic representation of the classical CSS magnetic
  state. (c) Ground state phase diagram of the model. The two
  horizontal dashed lines correspond to the results shown in
  Fig.~\ref{lc}. Transitions detected by critical level crossings are
  marked with "×," while transitions identified by ground state FS are
  indicated by "peak" symbols. The color and point-marks correspond to
  the real crossings and FS peaks shown in Fig.~\ref{lc}(c,d,e,f).}
\label{pd}
\end{figure}

\section{\label{phasediagram}  Ground state phase diagram}
The spin-$1/2$ $J_1$-$J_2$-$J_\chi$ model on a square lattice can be
expressed as follows,
\begin{eqnarray}
\nonumber
H=&&J_1\sum_{\langle i,j\rangle} \mathbf{S}_i \cdot  \mathbf{S}_j + J_2\sum_{\llangle i,j\rrangle}  \mathbf{S}_i \cdot  \mathbf{S}_j\\
\label{j1j2jx}
&&+J_\chi \sum_{\langle i,j,k \rangle_{\triangle}} ( \mathbf{S}_i \times  \mathbf{S}_j) \cdot  \mathbf{S}_k,
\end{eqnarray} 
where $\langle i,j\rangle$ and $\llangle i,j\rrangle$ denote the
nearest neighbor (NN) and the next nearest neighbor (NNN) pairs, while
$\langle i,j,k \rangle_{\triangle}$ signifies any smallest triangle
with its vertex sites $i, j, k$ arranged in a counterclockwise
order. These Hamiltonian terms are also highlighted in
Fig.~\ref{pd}(a). For the sake of simplicity, we fix $J_1=1$
throughout this paper and concentrate on the parameter space where
$J_2\in [0,1]$ and $J_{\chi}\in [0,1.4]$. The grand phase diagram is
presented in Fig.~\ref{pd}(c), which is the major result of this
study.

To determine the ground state phase diagram of this system, we employ
ED~\cite{Laflorencie04,Noack05,Weisse08,Lauchli11,Sandvik10ed} with
space group symmetry (rotation and translation to be specific), spin
inversion symmetry ($Z=\prod_i \sigma^x_i$), and the particle number
conservation symmetry ($S^z=\sum_is^z_i$). Each eigenstate is labeled
by a strict set of quantum numbers $(S,k_x,k_y,\phi_r)$, where $S$
represents the total spin, $k_x$, $k_y$ denote the momentum in the
$x$, $y$ directions, respectively, and $\phi_r$ signifies the phase
acquired when applying a $\pi/2$ lattice rotation given
$(k_x,k_y)=(0,0)$ or $(\pi,\pi)$. In cases where $k_x\neq k_y$, a
``$-$'' symbol is assigned to $\phi_r$, indicating that the rotation
symmetry is not applicable to this quantum sector. The spin inversion
quantum number $z$ on an even-site cluster is always $1$ for singlet
and $-1$ for triplet with total spin $S^z=0$, therefore not specified
again in the reminder of this paper.

We partition the two-parameter space into a mesh with steps of $\Delta
J_{\chi}=0.1$ and $\Delta J_2=0.025$. Fig.~\ref{lc} illustrates how
phase boundaries are determined for typical cuts in the two-parameter
space at $J_{\chi}=0.5$ and 1.0, as indicated by dashed lines in
Fig.~\ref{pd}(c), by scanning $J_2$ of a 32-site tilted cluster
(Fig.~\ref{pd}(a)). We analyze the ground states in the $(0,0,0,0)$
and $(0,0,0,\pi)$ sectors, displaying the two lowest energy levels in
each sector (Fig.~\ref{lc}(a,b)). Their separation between these
levels within each sector (energy gaps) indicates avoided level
crossings (Fig.~\ref{lc}(c,d)). Subsection~\ref{criticalcrossing}
details transitions identified by critical level crossings
(Fig.~\ref{lc}(c,d)).  Subsection~\ref{fs} discusses transitions using
peaks of FS (Fig.~\ref{lc}(e,f)) and energy second derivitives
(Fig.~\ref{lc}(g,h)). Their corresponding locations in the
two-parameter space are marked by ``$\times$'' sign and ``peak'' sign
along the dashed lines in Fig.~\ref{pd}(c) respectively.

\subsection{\label{criticalcrossing} Quantum Phase Transitions determined by Critical Level Crossings}
\begin{figure}
\centering
\includegraphics[width=0.45\textwidth]{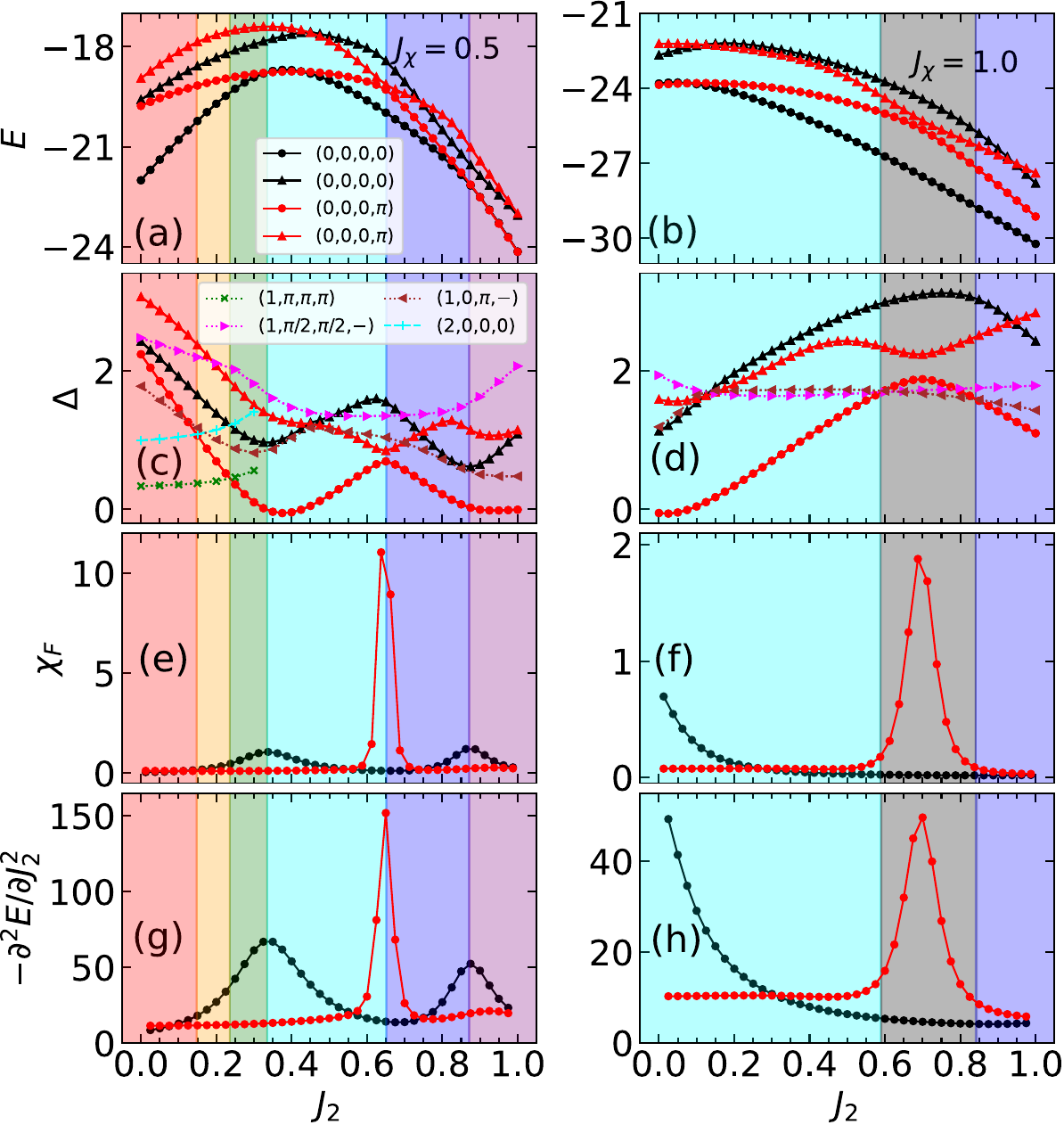}
\caption{(a, b) Two lowest energy levels in the
  $(S,k_x,k_y,\phi_r)=(0,0,0,0)$ and $(0,0,0,\pi)$ sectors for
  $J_{\chi}=0.5$ (a) and $1$ (b) as a function of $J_2$ for a 32-site
  tilted periodic cluster, as shown in Fig.~\ref{pd}(a). (c, d) Energy
  gaps relative to the lowest $(0,0,0,0)$ singlet for $J_{\chi}=0.5$
  (c) and $1$ (d). Gaps from relevant magnetic excitations, including
  the the $(1,\pi,\pi,\pi)$ triplet, $(2,0,0,0)$ quintuplet,
  $(1,0,\pi,-)$ triplet, and $(1,\pi/2,\pi/2,-)$ triplet, are also
  shown. In (c), the AFM-QSL transition is marked by a crossing
  between the $(2,0,0,0)$ quintuplet and the $(0,0,0,\pi)$ singlet,
  and the QSL-VBS transition by the crossing between the
  $(1,\pi,\pi,\pi)$ triplet and the $(0,0,0,\pi)$ singlet. In (d), the
  CSL-CSS and CSS-NSL transitions are marked by crossings between the
  $(1,\pi/2,\pi/2,-)$ triplet and the $(0,0,0,\pi)$ singlet. (e, f) FS
  for the lowest $(0,0,0,0)$ and $(0,0,0,\pi)$ singlets at
  $J_{\chi}=0.5$ (e) and $1$ (f), with energy second-order derivatives
  shown in (g) and (h). FS peaks align with energy derivative peaks
  and singlet energy gap dips within the same quantum sector. In (e),
  the left-black FS peak marks the VBS-CSL transition, while the
  right-black FS peak indicates the NSL-stripe transition. The red FS
  peak separates the CSL and NSL phases. Phase colors and boundary
  locations match the global phase diagram in Fig.~\ref{pd}(c).}
\label{lc}
\end{figure}

In this subsection, we discuss magnetic-to-non-magnetic phase
transitions in Fig.~\ref{pd}(c), indicated by ``$\times$'' sign
representing energy level crossings. The color and point-mark of these
``$\times$'' signs correspond to the real crossings shown in
Fig.~\ref{lc}(c,d). The three magnetic phases in the ground state
phase diagram are the well-known AFM and stripe phases, and the less
familiar chiral spin solid (CSS) phase.

The physics of CSS can be understood in the classical limit of
$J_{\chi}=\infty$ and $S\to \infty$. We follow the description of
Ref.~\cite{Rabson}. As shown in Fig.~\ref{pd}(b), classical CSS is
made of three interpenetrating decoupled antiferromagnetic
sublattices, the in-out aligned spins, the horizontally aligned spins,
and the vertically aligned spins. Translation symmetry can be restored
by superposition. Half of the smallest triangle classically persue the
largest negative value -1 for chiral interaction
$(\mathbf{S}_i\times\mathbf{S}_j) \cdot \mathbf{S}_k$ with $i,j,k$
arrange counterclockwisely. This state favors minimizing half of the
local term in the Hamiltonian $H_{\chi}=\sum_{\langle i,j,k
  \rangle_{\triangle}} ( \mathbf{S}_i \times \mathbf{S}_j) \cdot
\mathbf{S}_k$. One can check that this classical configuration has
antiferromagnetic order with the static structure factor peak at
$(\pm\pi,0)$, $(0,\pm\pi)$, and $(\pm\pi/2,\pm\pi/2)$. The name chiral
spin solid is respect to the fact that the local chiral order $\langle
\chi \rangle=\langle (\mathbf{S}_i\times\mathbf{S}_j) \cdot
\mathbf{S}_k\rangle$ is nonzero, while the state remains magnetically
ordered. At the smallest $S=1/2$, quantum fluctuation could possibly
destroy this magnetic order, as Ref.~\cite{Zhang24} has
indicated. However this is subject to further large scale numerical
examination sitting in the deep $J_{\chi}=\infty$ region.

Enumerating these magnetic phases, along with their low-energy
magnetic excitations, aids in selecting the correct highly symmetric
low-energy sectors for further analysis. For small $J_2$ and small
$J_{\chi}$, the ground state has AFM order, sitting in the $(0,0,0,0)$
quantum sector. Its gapless magnetic excitations include spin triplet
with quantum numbers $(1,\pi,\pi,\pi)$ and spin quintuplet with
quantum numbers $(2,0,0,0)$.

With large $J_2$ and small $J_{\chi}$, the ground states have
collinear/stripe order, and are two-fold degenerate. The states
feature antiferromagnetic spin correlation in one direction and
ferromagnetic spin correlation in the other. The positive and negative
superpositions of differently oriented stripe patterns form two
degenerate ground states in the $(0,0,0,0)$ and $(0,0,0,\pi)$ sectors
respectively. Its gapless magnetic excitations have quantum numbers
$(1,0,\pi,-)$ or $(1,\pi,0,-)$, revealing its stripiness.

In the limit of very large $J_{\chi}$, the ground state is likely
displaying CSS magnetic order~\cite{Rabson}. The gapless magnetic
excitations are expected to sit at $(k_x,k_y)=(\pm\pi/2,\pm\pi/2)$,
$(0,\pi)$, and $(\pi,0)$ in the Brillouin Zone, as in the spin-$1$
case~\cite{huang_coexistence_2022}.

For $J_{\chi}=0$, critical level crossings have already mapped out a
precise ground state phase
diagram~\cite{Wang18,Ferrari20,Nomura20,Wang22}. Between the
well-known antiferromagnetic (AFM) and collinear phases, there is a
gapless QSL and a columnar VBS phase. The AFM-to-QSL transition is
marked by a level crossing between a quintuplet magnetic excitation
and a low-energy singlet, which gradually evolves into the
quasi-degenerate VBS ground state as $J_2$ increases. The QSL-to-VBS
transition is identified by a crossing between a triplet magnetic
excitation and the VBS ground state. The VBS phase ends in a direct
first-order transition to the collinear phase at $J_2\approx 0.61$.

For $J_{\chi}>0$, we determine phase boundaries between the AFM and
gapless QSL, as well as between the gapless QSL and VBS, using the
same method as before. We plot energy gaps in Fig.~\ref{lc}(c,d) for
$J_{\chi}=0.5$ and $J_{\chi}=1$, relative to the ground state energy
in the $(0,0,0,0)$ sector, for the states shown in
Fig.~\ref{lc}(a,b). Gaps for relevant gapless magnetic excitations are
also included. The AFM-to-QSL transition is signaled by a crossing
between the $(2,0,0,0)$ quintuplet and the $(0,0,0,\pi)$ singlet,
while the QSL-to-VBS transition is marked by a crossing between the
$(1,\pi,\pi,\pi)$ triplet and the $(0,0,0,\pi)$ singlet. By simulating
across various $J_{\chi}$, we identify two critical lines in the
global phase diagram in Fig.~\ref{pd}(c), delineating the AFM-to-QSL
and QSL-to-VBS transitions.

\begin{figure}[t]
  \includegraphics[width=0.5\textwidth]{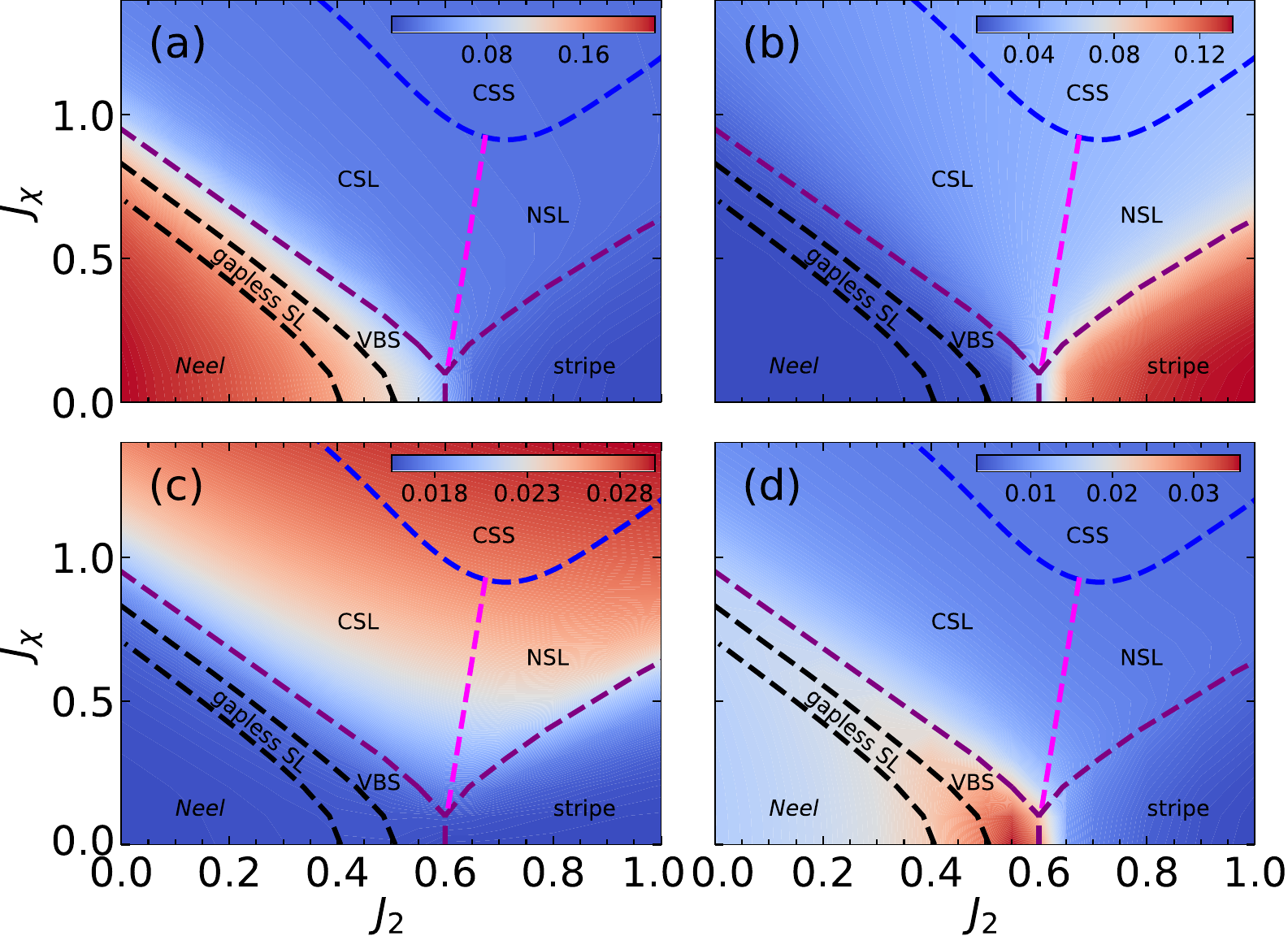}
  \caption{Various local order parameters in color contour computed
    using the ground state in the $(0,0,0,0)$ sector for an $N=32$
    periodic cluster. They are staggered magnetization square
    $m^2(\pi,\pi)$ (a), collinear magnetization square $m^2(0,\pi)$
    (b), the CSS magnetization square $m^2(\pi/2,\pi/2)$ (c),
    and dimer order parameter square $d_{\hat{x}}^2(\pi,0)$ for the
    VBS phase (d). The color contour lines match the boundaries of
    individual local ordered phases. Phase boundaries are shown in
    white dashed lines.}
  \label{order}
\end{figure}

In Fig.~\ref{lc}(d) at $J_{\chi}=1$, two magnetic triplet excitations,
$(1,\pi/2,\pi/2,-)$ and $(1,0,\pi,-)$ drop below all singlet
excitations, suggesting a non-degenerate ground state with likely
gapless magnetic excitations. The strong static structure factor at
$(\pi/2,\pi/2)$ in Fig~\ref{order}(c) indicates a CSS phase, similar
to that found in the spin-$1$ $J_1$-$J_2$-$J_{\chi}$ model at large
$J_{\chi}$~\cite{huang_coexistence_2022}. Zhang et al. observed a
vanishing magnetic order for $J_{\chi}=J_2=1$~\cite{Zhang24}, which
aligns with the NSL phase in our global phase diagram, consistent with
our findings.

We define the region in parameter space where magnetic excitations are
the lowest as the CSS phase, as shown in Fig.~\ref{pd}(c). The
transition into this phase is indicated by a critical level crossing
between magnetic and non-magnetic states, specifically between the
$(1,\pi/2,\pi/2,-)$ triplet and the $(0,0,0,\pi)$ singlet.  This
crossing determines the left and right boundaries of the CSS phase for
a given value of $J_{\chi}$.

\subsection{\label{fs} Quantum Phase Transitions determined by Fidelity Susceptibility and avoided ground state level crossing}

Other phase boundaries appearing in Fig.~\ref{pd}(c) are determined by
peaks in the ground state FS, and the peaks consistently arising in
the second-order energy derivatives.

FS is commonly used to detect phase transition accompanied by avoided
ground state level crossing~\cite{GuFidelity10,You11,Wang10}. Its
definition is as following
\begin{equation}
\chi_F(g)= {\lim_{ \delta g \to 0}} \frac{-2\ln \vert \langle\phi(g+\delta g)|\phi(g)\rangle\vert }{ (\delta g)^2},
\end{equation}
where $\phi(g)$ is the lowest energy state of a given quantum sector
at parameter $g$. The energy second-order derivative is
straightforwardly defined as
\begin{equation}
\frac{ \partial^2 E(g)}{\partial g^2}=\frac{ \big(E(g+\delta
  g)+E(g-\delta g)-2E(g)\big)}{(\delta g)^2}.
\end{equation}
In the gapped non-magnetic phases—VBS, CSL, and NSL—the ground state
degeneracy spans the $(0, 0, 0, 0)$ and $(0, 0, 0, \pi)$ sectors.
Fig.~\ref{lc}(e,f) show the ground state FS, while Fig.~\ref{lc}(g,h)
display the second-order energy derivative for each sector.

At $J_{\chi}=0.5$, as shown in Fig.~\ref{lc}(a,c,e,g), we observe
three FS peaks (Fig.~\ref{lc}(e))—two in black for the $(0,0,0,0)$
sector and one in red for the $(0,0,0,\pi)$ sector —coinciding with
the locations of the peaks in energy derivatives (Fig.~\ref{lc}(g))
and the dips in the singlet energy gap in the corresponding sector
(Fig.~\ref{lc}(c)). The exact correspondance showcase the fact the
energy second-order derivative is in proportion to the FS by
wavefunction purtabation theory~\cite{GuFidelity10}. The left-black FS
peak marks the transition from the columnar VBS to the CSL phase,
while the right-black FS peak indicates the transition from the NSL to
the stripe magnetic phase. The prominent red FS peak separates the CSL
and NSL phases. The continuity of the energy second-order derivative
(in Fig.~\ref{lc}(g)) suggests the nature of the three phase
transitions is likely to be continuous.

At $J_{\chi}=1$, as shown in Fig.~\ref{lc}(b,d,f,h), we observe a
partial left-black FS peak and a full red FS peak. The missing
right-black peak and the incomplete left-black FS peak in
Fig.~\ref{lc}(f), compared to Fig.~\ref{lc}(e), occur because the
intersections of the dashed horizontal line at $J_{\chi}=1$ with the
VBS-CSL and NSL-stripe phase boundaries lie outside the range $J_2\in
[0,1]$, as shown in Fig.~\ref{pd}(c).

At $J_{\chi}=1$, the red FS peak in Fig.~\ref{lc}(f) is not linked to
the CSL-NSL phase transition, as the unique ground state here is the
CSS. Consequently, the dashed horizontal line at $J_{\chi}=1$ in
Fig.~\ref{pd}(c) does not intersect the CSL-NSL phase boundary.

In summary, the three FS peaks in the relevant region define the phase
boundaries between the VBS-CSL, CSL-NSL, and NSL-stripe phases, as
shown in Fig.~\ref{pd}(c).

\subsection{\label{phasedigram}Full Ground State Phase Diagram}

In Fig.~\ref{pd}(c), we present the ground state phase diagram of the
$J_1$-$J_2$-$J_{\chi}$ model, summarizing all findings from critical
level crossings and ground state FS analysis. Aside from a direct
VBS-stripe transition known from the $J_1$-$J_2$ model, all discussed
critical transitions appear continuous. This is supported by two
observations: the absence of discontinuities in the second-order
energy derivative, and the smooth energy behavior across transition
points in the other degenerate ground states showing no FS peak.

We identify three new phases—the CSL, NSL, and magnetic CSS phases—in
addition to the four known phases in the $J_1$-$J_2$ model
(antiferromagnetic, gapless QSL, columnar VBS, and collinear
phases). Before exploring the topological nature of the two new
non-magnetic phases, we first examine their local order parameters.

\begin{figure}
  \includegraphics[width=0.4\textwidth]{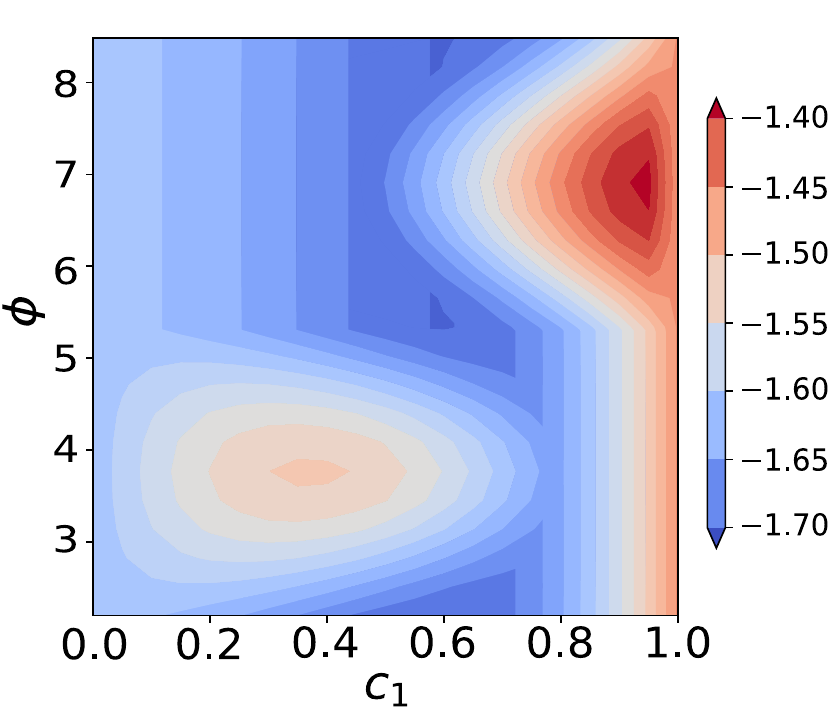}
  \caption{The negative of EE defined as
    $-S=\text{log}\text{Tr}(\rho_A^2)$ for superimposed state
    $|\psi^l_v\rangle$ via $|\xi_1\rangle$ and $|\xi_2\rangle$ as a
    function of $c_1,\phi$ in Eq.~\ref{mesdef} at $J_\chi=0.5$ and
    $J_2=0.5$ on a $N=32$ cluster, where $\rho_A$ is the reduced
    density matrix of a half of the system defined following a
    diagonal cut (horizontally or vertically in Fig.~\ref{pd}(a)) that
    bipartite the system into equal sized two. $|\xi_1\rangle$ and
    $|\xi_2\rangle$ are two topological degenerate states lives in
    (0,0,0,0) and (0,0,0,$\pi$) sector respectively.}
  \label{mes}
\end{figure}

\section{\label{localorders} Local order parameters}
To describe the magnetic and valence bond ordered phases, we define
the following local order parameters. The magnetization square at
momentum $\vec{k}=(k_x,k_y)$ is defined as
\begin{equation}
	m^2(\vec{k})=\frac{1}{N^2}\sum_{lm}e^{i\phi_{lm}}\langle \mathbf{S}_l \cdot \mathbf{S}_m \rangle,
\end{equation}
where $\phi_{lm}=\vec{k}\cdot(\vec{r}_l-\vec{r}_m)$. For the AFM
phase, $m^2(\vec{k})$ peaks at momentum $\vec{k}=(\pi,\pi)$, while for
the collinear phase it peaks at momentum $(0,\pi)$ and $(\pi,0)$. In
the CSS phase, there are 4 main peaks at $(\pm\pi,0)$ and
$(0,\pm\pi)$, along with four satellite peaks at
$(\pm\pi/2,\pm\pi/2)$~\cite{huang_coexistence_2022}. We use
$(q_x,q_y)=(\pi/2,\pi/2)$ as a distinguishing momentum for the CSS
phase to differentiate it from the collinear phase.

The dimer order parameter is defined as
\begin{equation}
	d_{\mathbf{\alpha}}^2(\vec{k})=\frac{1}{N^2}\sum_{lm}\theta^{\alpha}_{lm}\langle D^{\mathbf{\alpha}}_l D^{\mathbf{\alpha}}_m\rangle,
\end{equation}
where $\alpha=x,y$, $\theta^{\alpha}_{lm}=(\pm 1)^{(r_l^{\alpha}-r_m^{\alpha})}$, $D^{x}_m=\mathbf{S}_{\vec{r}_m}\cdot \mathbf{S}_{\vec{r}_m+\vec{e}_x}$, and  $D^{y}_m=\mathbf{S}_{\vec{r}_m}\cdot \mathbf{S}_{\vec{r}_m+\vec{e}_y}$.

All four local order parameters are computed using the ground state in
the $(0,0,0,0)$ sector, and their color contour plots are illustrated
in Fig.~\ref{order}. We observe that the contours of local order
parameters are roughly consistent with various phase boundaries of the
ground state phase diagram.

\section{\label{csl}Topological nature of the CSL phase}
The CSL state is a topological state characterized by semion
excitations in the bulk and gapless edge excitations described by an
$\rm{SU}(2)_1$ Wess-Zumino-Witten conformal field
theory~\cite{Wen91}. Semions, which are spinons in a chiral spin
liquid, carry spin-1/2 but no charge. When a semion moves along a
closed path, it generates a flux phase depending on the number of
enclosed semions. Unlike bosons or fermions, exchanging a pair of
semions results in a phase shift of $\pi/2$, hence the name
``semion''. Different ground states arise by inserting different
fluxes into a non-contractable loop~\cite{WenWilczekZee}.

To uncover the topological nature of the CSL phase, modular
$\mathcal{S}$ matrix is often used to extract nontrivial statistics of
quasiparticle~\cite{YZhang12}. We compute the $\mathcal{S}$ matrix,
employing two degenerate singlets obtained using ED on a 32-site
torus. It is defined as the overlap matrix between the minimally
entangled states (MES) along cuts in $x$ and $y$ directions. Here the
MES along a cut in $x$ ($y$) direction is numerically optimized by
minimizing its entanglement entropy (EE) with respect to superimposing
parameters $\phi$ and $c_1$ (defined below) of the two quasi
degenerate ground states.

\begin{figure}
\centering \includegraphics[width=0.35\textwidth]{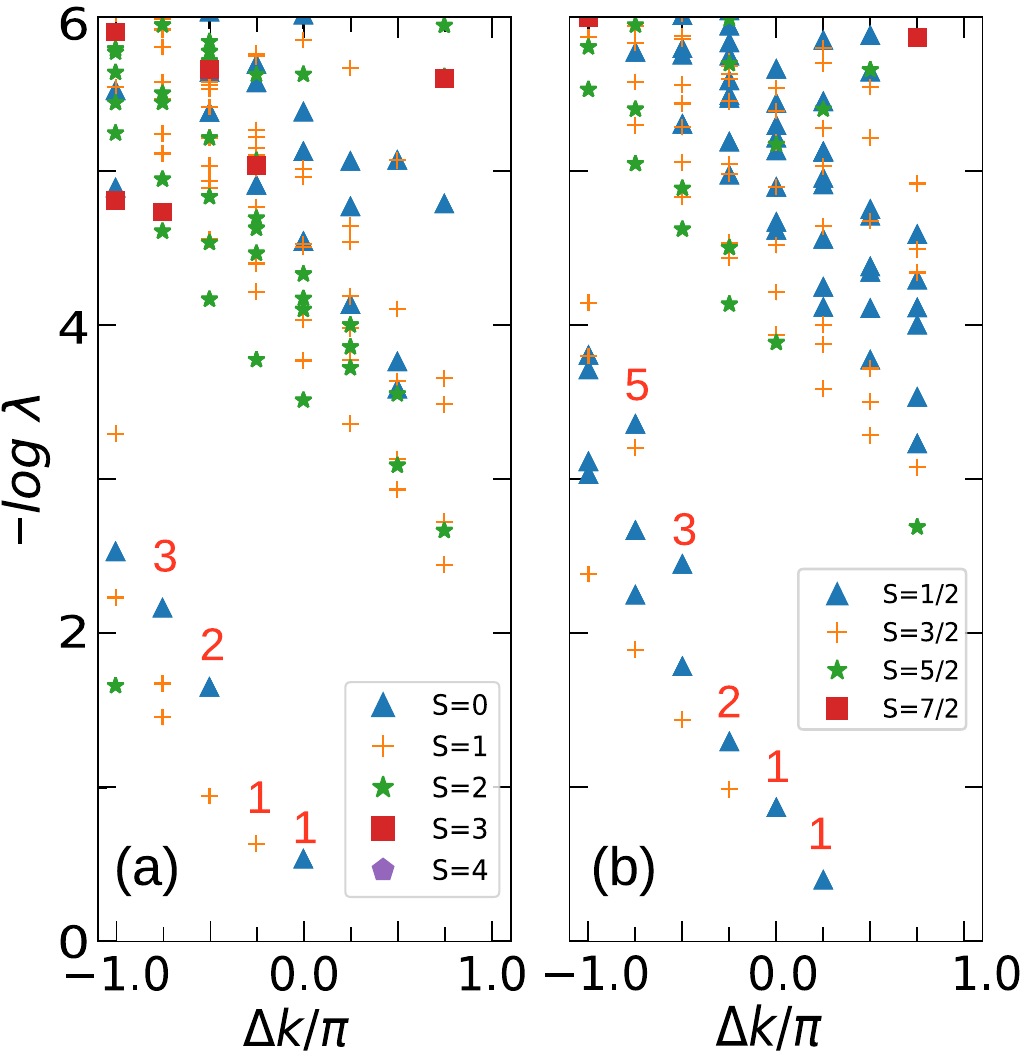}
\caption{Bipartite ES for the CSL ground states in even (a) and odd
  (b) sectors of a $2L \times L$ cylinder at $L=8$, $J_2=0.5$ and
  $J_\chi=0.5$. Each eigenvalues of the reduced density matrix
  $\rho_{L}$ is associated with a total spin quantum number $S$ and a
  momentum quantum number $k_y$, which is obtained from the phase
  difference of individual eigenvectors of $\rho_L$ before and after
  the action of translation operator on the left half of the
  wavefunction.}
\label{es}
\end{figure}

Referring to a winding direction ($l=x,y$) on a torus, we name the two
topological ground states as $|\psi^l_v\rangle$ and $|\psi^l_s\rangle$
respectively ($v$ for vacuum and $s$ for semion). ED calculation does
not produce $|\psi^l_v\rangle$ and $|\psi^l_s\rangle$ directly. The
quasi degenerate eigenstates obtained via ED, say $|\xi_1\rangle$ from
the $(0,0,0,0)$ sector and $|\xi_2\rangle$ from the $(0,0,0,\pi)$
sector, are combinations of $|\psi^l_v\rangle$ and $|\psi^l_s\rangle$,
as
\begin{eqnarray}
  \nonumber
  |\psi^l_v\rangle&=&c_1|\xi_1\rangle +c_2e^{i\phi}|\xi_2\rangle,\\
  \label{mesdef}
  |\psi^l_s\rangle&=&c_2|\xi_1\rangle -c_1e^{i\phi}|\xi_2\rangle.  
\end{eqnarray}
By definition, minimizing the EE of $|\psi^l_v\rangle$ can settle the
two parameters $c_1,\phi$ ($c_2\equiv\sqrt{1-c_1^2}$) for a given
$l=x,y$. Once having $|\psi^l_v\rangle$ and $|\psi^l_s\rangle$, we
define two two-component vectors
$|\Psi^l\rangle=\{|\psi_v^l\rangle,|\psi_s^l\rangle\}$. The modular
$\mathcal{S}$ matrix is formally written as
$\mathcal{S}=\langle\Psi^x|\Psi^y\rangle$.

Indeed, through the minimization procedure of $|\Psi^x\rangle$ we find
two EE minima (refer to $|\psi^x_v\rangle$, $|\psi^x_s\rangle$
respectively), as in Fig.~\ref{mes}(a). Their relative positions on
$c_1$ axis square-sum to 1, and their superimposing phase difference
is $\pi$. For the other pair of MES $|\Psi^y\rangle$, we take
advantage of the rotation symmetry, and find
$|\psi^{y}_v\rangle=R_{\pi/2}|\psi^{x}_v\rangle$ and
$|\psi^{y}_s\rangle=R_{\pi/2}|\psi^{x}_s\rangle$. We thus identify the
$\mathcal{S}$ matrix within the CSL phase at parameters $J_2=0.5$ and
$J_{\chi}=0.5$ as
\begin{equation}
  \label{modulars}
\mathcal{S}=0.750
\begin{pmatrix}
0.968 &  0.890\\
0.890 & -1.038
\end{pmatrix}
\approx
\frac{1}{\sqrt{2}}
\begin{pmatrix}
1 &  1\\
1 & -1
\end{pmatrix}.
\end{equation}
This result confirms the expected semion mutual statistics with
total quantum dimension of $\sqrt{2}$~\cite{zhu_minimal_2013}.

The topological fractional quantum Hall state has gapless edge states. Li and Haldane~\cite{LiHaldane} proposed that the entanglement spectrum of the reduced density matrix of ground states in time-reversal-breaking topological phases reflects the counting of edge modes. This is seen when the ground state is divided into two spatial regions, and one region is traced out. By numerically obtaining the topological ground states on a cylinder and analyzing their entanglement spectra, we can infer the counting of their edge modes as if the system had a physical edge.

The chiral spin liquid (CSL) phase has two quasi-degenerate ground states on a cylinder with an exponentially small energy difference. In our model, the ground state in the vacuum sector has the lowest energy and is easiest to find using brute-force DMRG. The ground state in the semion sector lies slightly higher. To target the semion sector directly, one can thread a $\pi$ flux through the cylinder, which effectively pumps a spin-1/2 semion through it, leaving net spin-1/2 semions on both sides. This approach is equivalent to removing one spin from each boundary of the cylinder, which also introduces semions. The entanglement spectra of the semionic ground state reveals odd half-integer boundary edge modes.  

To investigate edge physics, we study the CSL phase on a $2L\times L$ cylinder with perimeter $L=4,6,8$ using an $\rm{SU}(2)$ symmetric DMRG algorithm~\cite{Wbaum12}, with open boundaries along the $x$-axis and periodic boundaries along the $y$-axis. By imagining a vertical cut through the cylinder, the ground state converges to the vacuum sector $|\psi_v\rangle$ when there is an even number of sites on both sides. If one spin is removed from each side, the ground state instead converges to the semion sector $|\psi_s\rangle$. Fig.~\ref{es} shows the bipartite entanglement spectra for $|\psi_v\rangle$ (a) and $|\psi_s\rangle$ (b) at $J_2=0.5$ and $J_{\chi}=0.5$ with $L=8$. Each eigenvalue of the reduced density matrix $\rho_L$ corresponds to a total spin quantum number $S$ for the left half and a momentum quantum number $k_y$  from the translation operator acting on $\rho_L$'s eigenvector. Both entanglement spectra exhibit a degeneracy pattern (1, 1, 2, 3, 5, 8,$\cdots$), consistent with the Kac-Moody tower of descendants of the identity and spin-$1/2$ primary fields~\cite{kacmoodybook}.

\begin{figure}
  \includegraphics[width=0.5\textwidth]{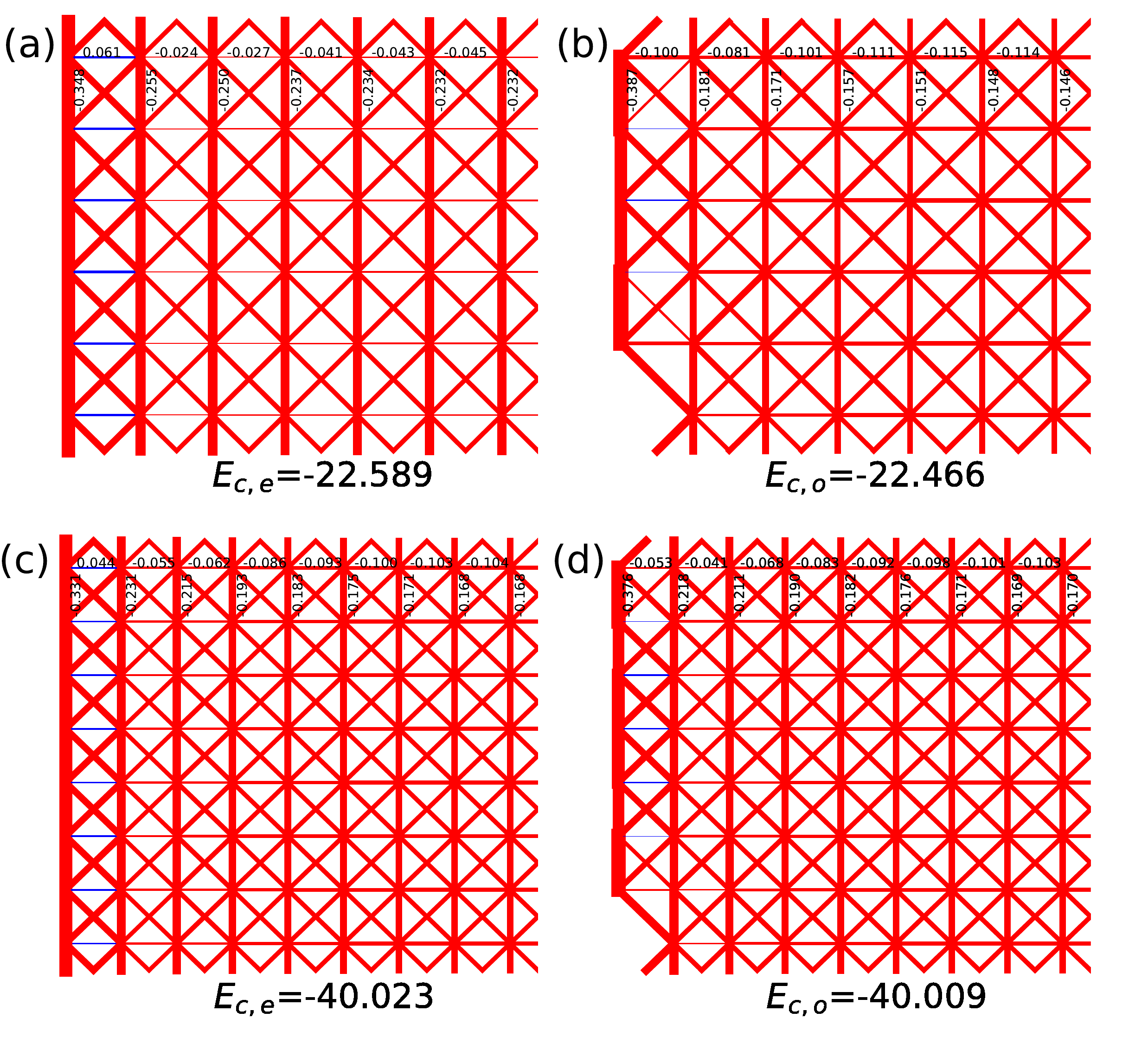}
  \caption{The bond energy landscape (of the left symmetric half) for
    even and odd sectors of a $2L\times L$ cylinder with $L=6,8$ at
    $J_2=0.7$ and $J_{\chi}=0.5$, within the NSL phase. The thickness
    of the line is proportional to the absolute value of individual
    bond energy, which is explicitly shown. Blue color represents
    positive bond energy, and red color represents negative bond
    energy. The bottom measurement $E_{c,e/o}$ shows the total energy
    of the center $L\times L$ area, demonstrating an almost perfect
    quasi-degeneracy in ground state energy between even and odd
    sectors.}
  \label{bond_energy}
\end{figure}

\begin{figure}
  \includegraphics[width=0.5\textwidth]{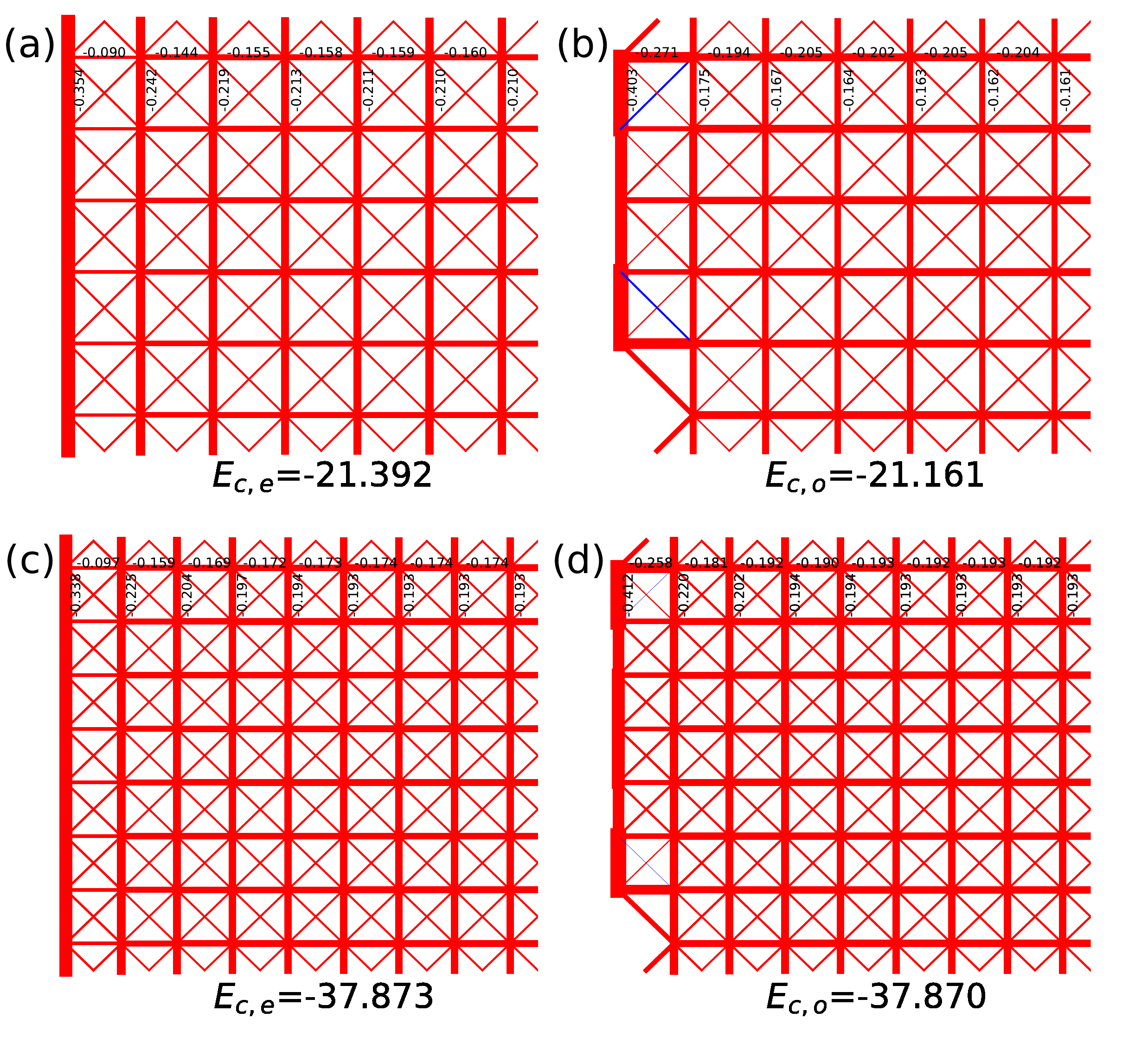}
  \caption{The bond energy landscape (of the left symmetric half) for
    the identity and spinon sectors of a $2L\times L$ cylinder with
    $L=6,8$ at $J_2=0.5$ and $J_{\chi}=0.5$, within the CSL phase. The
    line thickness and colors are similarly defined as in
    Fig.~\ref{bond_energy}.}
  \label{bond_energy_csl}
\end{figure}

\section{\label{nsl}Nature of the NSL phase}
To elucidate the characteristics of the NSL phase, we analyze the bond
energy landscape across variously sized cylinders, examining both even
and odd sectors, to discern any anisotropic trends in the
thermodynamic limit. Our investigation reveals a pronounced bond
anisotropy within the NSL phase. For clarity, we introduce
$E_{x/y,e/o}$ as the bond energy obtained by evaluating
$\mathbf{S}_i\cdot \mathbf{S}_j$ for a nearest-neighbor pair $\langle
i,j\rangle$ aligned along the $x/y$ direction on an even/odd cylinder,
respectively. The bond anisotropy is then quantified by computing the
energy differences in the $x$ and $y$ directions,
\begin{eqnarray}
  \nonumber
  \Delta E_e&=&E_{x,e}-E_{y,e},\\
  \label{bonddiff}
  \Delta E_o&=&E_{x,o}-E_{y,o}.
\end{eqnarray}
These values are summarized in Table~\ref{bond_anisotropy} and can
also be visually inspected in Fig.~\ref{bond_energy}. The measure of
nematicity from both even and odd sectors converges as the system
width $L$ increases, with its value particularly amplified on a size
$L = 8$ odd cylinder. Conversely, in the CSL phase, while the bond
energy anisotropy from even and odd sectors also converges, their
values diminish, as evidenced in Table~\ref{bond_anisotropy_csl} and
Fig.~\ref{bond_energy_csl}.

\begin{table}
  \begin{center}
    \caption{\label{bond_anisotropy} Bond energy measured in the $x/y$
      direction on an even/odd cylinder with width $L$, along with
      their directional anisotropy as defined in Eq.~\ref{bonddiff},
      is presented for the NSL phase at $J_2=0.7$ and $J_{\chi}=0.5$.}
    \vskip2mm
 \begin{tabular}{|l|l|l|l|l|l|l|}
\hline
~$L$~ & ~$E_{y,e}$~ & ~$E_{x,e}$~ & ~$E_{y,o}$~ & ~$E_{x,o}$~ & ~$\Delta E_{e}$~ & ~$\Delta E_{o}$~
\tabularnewline \hline
~6~ & ~-0.232~ & ~-0.045~ & ~-0.146~ & ~-0.117~ & ~0.187~ & ~0.029~
\tabularnewline \hline
~8~ & ~-0.168~ & ~-0.104~ & ~-0.170~ & ~-0.103~ & ~0.064~ & ~0.067~
\tabularnewline \hline 
\end{tabular}
\end{center}
\end{table}

\begin{table}
  \begin{center}
    \caption{\label{bond_anisotropy_csl} Bond energy measured in the
      $x/y$ direction on an even/odd cylinder with width $L$, along
      with their directional anisotropy as defined in
      Eq.~\ref{bonddiff}, is presented for the CSL phase at $J_2=0.5$
      and $J_{\chi}=0.5$.}
    \vskip2mm
 \begin{tabular}{|l|l|l|l|l|l|l|}
\hline
~$L$~ & ~$E_{y,e}$~ & ~$E_{x,e}$~ & ~$E_{y,o}$~ & ~$E_{x,o}$~ & ~$\Delta E_{e}$~ & ~$\Delta E_{o}$~
\tabularnewline \hline
~6~ & ~-0.210~ & ~-0.160~ & ~-0.161~ & ~-0.204~ & ~0.050~ & ~-0.043~
\tabularnewline \hline
~8~ & ~-0.193~ & ~-0.174~ & ~-0.193~ & ~-0.192~ & ~0.019~ & ~~0.001~
\tabularnewline \hline 
\end{tabular}
\end{center}
\end{table}

We further evaluate the total energy of the central $L\times L$ area
within the $2L\times L$ even and odd cylinders for both the NSL and
CSL phases, denoted as $E_{c,e/o}$ in Fig.~\ref{bond_energy} and
Fig.~\ref{bond_energy_csl}, respectively. In the NSL phase, their
values at $L = 6$ are $E_{c,e} = -22.589$ and $E_{c,o} = -22.466$ for
even and odd sectors, respectively, while at $L = 8$, $E_{c,e} =
-40.023$ and $E_{c,o} = -40.009$ for even and odd, respectively. This
illustrates that the energy gap between even and odd sectors of the
NSL on a cylinder diminishes exponentially with $L$. A similar
behavior is observed in the CSL phase, as depicted in
Fig.~\ref{bond_energy_csl}. We anticipate the ground state manifold to
double ({\it i.e.} becoming quadruple) on a torus for the NSL phase,
since the direction of nematicity can appear in either the $x$ or $y$
direction.

\begin{figure}[t]
  \includegraphics[width=0.35\textwidth]{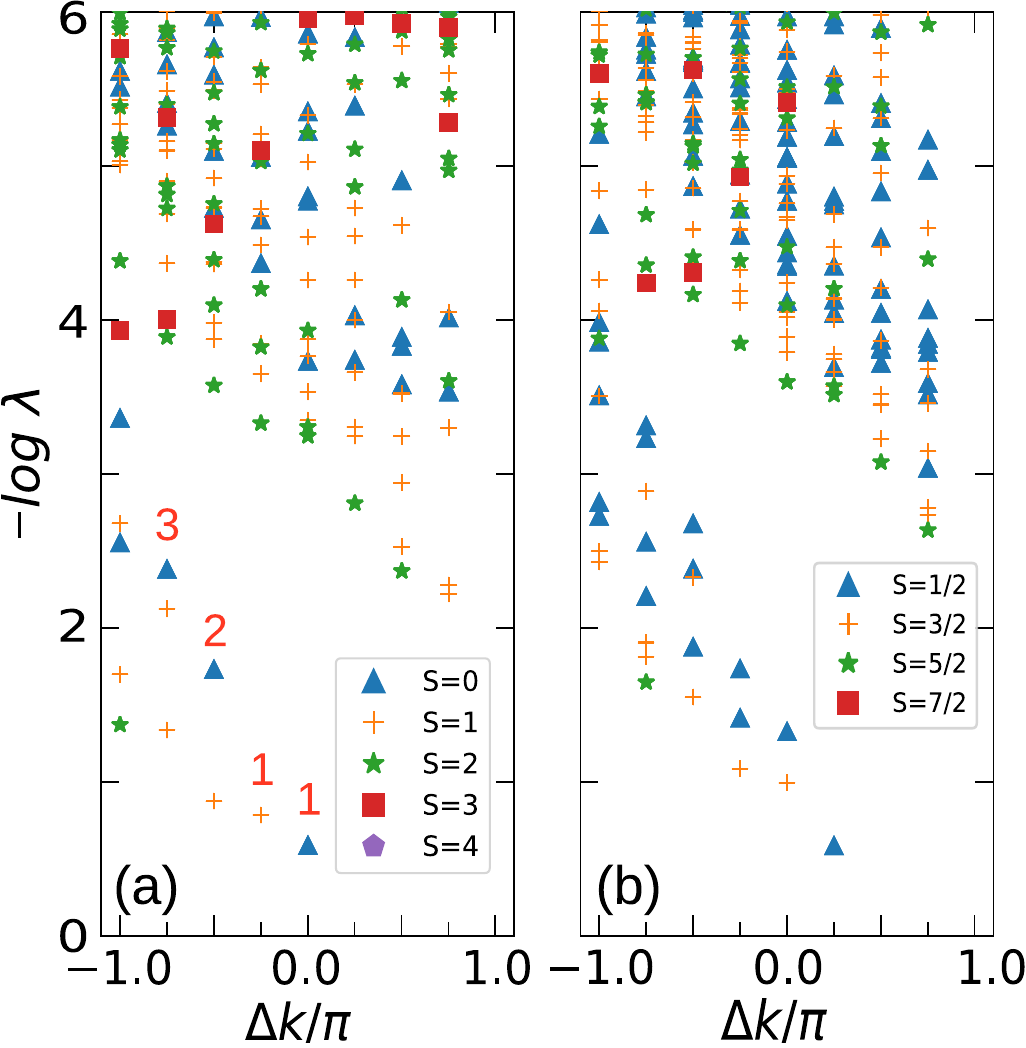}
  \caption{Bipartite ES for the NSL ground states in even (a) and odd
    (b) sectors of a $2L \times L$ cylinder at $L=8$, $J_2=0.7$ and
    $J_\chi=0.5$. The odd sector has an incomplete degeneracy sequence
    in comparison with that appears in Fig~\ref{es} for the CSL.}
  \label{es_x}
\end{figure}

To contrast the NSL with the CSL phase, we present the edge ES on a
cylinder with perimeter of $L=8$, measured at $J_2=0.7$,
$J_{\chi}=0.5$ in Fig.~\ref{es_x}. The ES in the even sector exhibits
the same chiral edge modes as those observed in the CSL. However, the
odd sector behaves differently compared to its CSL counterpart, as
shown in Fig.~\ref{es}(b), providing additional evidence of being a
new phase.

\begin{figure}[t]
  \includegraphics[width=0.3\textwidth]{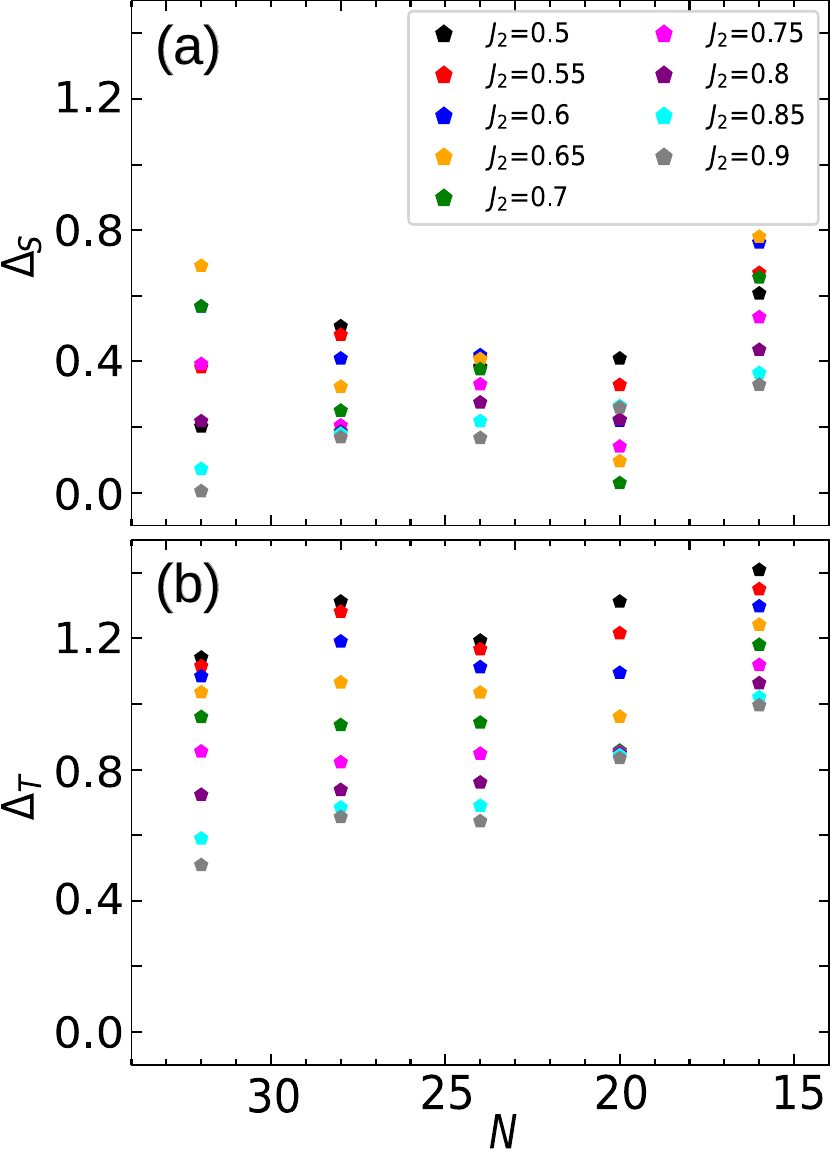}
  \caption{The gap splitting between the two lowest singlets among all
    sectors (a), and the triplet gap between the $(1,0,\pi,-)$ stripe
    triplet and the lowest singlet among all sectors (b) at
    $J_{\chi}=0.5$, for various $J_2$ on a series of finite size
    periodical clusters with $N=16,20,24,28,32$. The $x$ axis takes
    the form $N$ in reverse order to avoid any potential misleading
    associated with finite size extrapolation.}
  \label{gaps}
\end{figure}

On a torus, we define $\Delta_S$ as the energy splitting between the
lowest two singlets across all sectors, and $\Delta_T$ as the gap
between the $(1,0,\pi,-)$ triplet and the lowest singlet. We present
their $\Delta_S$ and $\Delta_T$ values in Fig.~\ref{gaps} for
$J_{\chi} = 0.5$ and $J_2 \in [0.5, 0.9]$ with step $\Delta J_2 =
0.05$, considering periodic clusters of sizes $N=16,20,24,28,32$. We
observe that all $\Delta_S$ consistently resides below the
corresponding $\Delta_T$, and the triplet gaps tend to stabilize as
$J_2$ deviates from the transition coupling $J_{2}^c\approx
0.87$. This reaffirms the ground state degeneracy and the persistence
of the triplet gap within the CSL and NSL phases. We closely examine
the spectrum of the 32-site torus within the NSL region, aiming to
identify all four ground states from Fig.~\ref{lc}(c). We notice that
the two singlets of higher energy in each sector are positioned
above the lowest triplet $(1,0,\pi,-)$ state. However, this is
attributed to finite size effects. We anticipate the four shown
singlets will eventually descend below the triplet gap as the energy
splittings between them become exponentially small with increasing
size $L$.

\section{\label{conclusion}Conclusions}

We investigate the ground state phase diagram of the spin-$1/2$
$J_1$-$J_2$-$J_{\chi}$ model on a square lattice. Our analysis reveals
a complex two-parameter phase diagram. For small and fixed
$J_{\chi}\in [0:0.1]$, the one-parameter phase diagram of the
$J_1$-$J_2$ model~\cite{Wang18,Ferrari20,Nomura20} naturally extends
to $J_{\chi}\neq 0$, encompassing four well-known phases, though these
are not the primary focus of our study. At intermediate $J_{\chi}$
values, a topological chiral spin liquid (CSL) state and a nematic
spin liquid (NSL) state emerge concurrently, situated between a
valence bond solid (VBS) and a collinear magnetic state. Further, a
chiral spin solid (CSS) magnetic state envelops the two spin liquid
phases at higher $J_{\chi}$ values. The boundaries between these
phases are determined using critical level crossing arguments between
magnetic and non-magnetic phases, or through peaks in fidelity
susceptibility (FS) accompanied by avoided ground state level
crossings.

Our phase diagram for the region where the CSL exists closely aligns
with an earlier ED study on a $4\times 5$ cluster~\cite{Nielsen13},
although the authors of that study overlooked the NSL phase and the
CSS magnetic phase. Within the proposed NSL phase, level spectroscopy
reveals a quasi four-fold ground state degeneracy on a torus,
consistent with rotation symmetry breaking in the bond energy
landscape and the two-fold ground state manifold on a cylinder. When
comparing entanglement spectra (ES) of the NSL and the CSL on a
cylinder, we observe an unrecognized edge behavior in the odd
topological sector, whose counting does not match any conformal field
theories known to us. Another possibility could be that there is no
edge mode at all for the NSL, and those appearing in the ES are a
finite size effect. This question will be addressed in future
studies. Our findings offer a concrete example of a nematic spin
liquid state arising from a realistic lattice Hamiltonian.

\section*{Acknowledgements}
Note added. Recently, we became aware of an article by zhang el
al.~\cite{Zhang24}, who studied the same model by DMRG. Both of the
two works consistently show the AFM, stripe, CSL, and the NSL (named
disordered phase in their study) phases. However their DMRG finite
size extrapolation doesn't support the surviving of the CSS order in
the thermodynamic limit. While our work doesn't focus on the finite
size extrapolation of the CSS order though. This discrepency shall
deserve further inceasing of the cylinder size in the furture.

We would like to thank S.-S. Gong for discussion and manuscript
sharing, and C. Xu for pointing to us the possibility of a NSL phase,
and W. Zhu, G. M. Zhang and H. H. Tu for enlightening
discussion. Z.L. is supported by the National Key Research and
Development Program of China (2020YFA0309200). L.W. is supported by
the National Natural Science Foundation of China, Grants
No.~NSFC-12374150 and No.~NSFC-11874080 .

\end{document}